\RequirePackage{lineno}
\documentclass[aps,prc,reprint,prcsuperscriptaddress,nofootinbib,11pt]{revtex4}
\usepackage{epsfig}
\usepackage{graphicx}% Include figure files
\usepackage{dcolumn}% Align table columns on decimal point
\usepackage{bm}% bold math
\usepackage{color}
\usepackage{url}
\usepackage{multirow}
\usepackage{enumitem}

\newcommand {\dAu}	{$d$+Au}
\newcommand {\pAu}	{$p$+Au}

\newcommand {\dndeta}	{dN_{ch}/d\eta}

\newcommand {\gOS}	{\gamma_{\rm OS}}
\newcommand {\gSS}	{\gamma_{\rm SS}}
\newcommand {\dg}	{\Delta\gamma}
\newcommand {\dgscale}	{\Delta\gamma_{\rm scaled}}

\begin{document}
%% Title, authors and addresses
%% use the tnoteref command within \title for footnotes;
%% use the tnotetext command for theassociated footnote;
%% use the fnref command within \author or \address for footnotes;
%% use the fntext command for theassociated footnote;
%% use the corref command within \author for corresponding author footnotes;
%% use the cortext command for theassociated footnote;
%% use the ead command for the email address,
%% and the form \ead[url] for the home page:
%% \title{Title\tnoteref{label1}}
%% \tnotetext[label1]{}
%% \author{Name\corref{cor1}\fnref{label2}}
%% \ead{email address}
%% \ead[url]{home page}
%% \fntext[label2]{}
%% \cortext[cor1]{}
%% \address{Address\fnref{label3}}
%% \fntext[label3]{}

%% use optional labels to link authors explicitly to addresses:
%% \author[label1,label2]{}
%% \address[label1]{}
%% \address[label2]{}

\title{HIJING can describe the anisotropy-scaled charge-dependent correlations at the Relativistic Heavy Ion Collider}
%%%%%%%%%%%%%%%%%%%
\author{Jie Zhao}
\address{Department of Physics and Astronomy, Purdue University, West Lafayette, IN 47907, USA}

\author{Yicheng Feng}
\address{Department of Physics and Astronomy, Purdue University, West Lafayette, IN 47907, USA}

\author{Hanlin Li}
\address{Hubei Province Key Laboratory of Systems Science in Metallurgical Process, Wuhan University of Science and Technology, Wuhan, Hubei 430081, China}

\author{Fuqiang Wang}
\email{fqwang@purdue.edu}
\address{Department of Physics and Astronomy, Purdue University, West Lafayette, IN 47907, USA}
\address{School of Science, Huzhou University, Huzhou, Zhejiang 313000, China}

\date{\today}

\begin{abstract}
The experimentally measured charge-depdendent correlations in heavy ion collisions have been suggested as a signature of the chiral magenetic effect (CME). 
Early model studies could not reproduce the measurement. 
For example, the Hijing model yielded far smaller magnitude for the charge-dependent correlation than observed in data.
This led to the conclusion that the CME had to be invoked to explain the observed correlations in heavy ion collisions.
In this paper we show that this conclusion of the CME interpretation is premature. We show that the reason that Hijing predicts a far smaller correlation than data is because the elliptic anisotrpy ($v_{2}$) parameter in Hijing is too small. When properly scaled, the Hijing model can reproduce in entirety the measured correlations. 
We also employ the AMPT model, which has a large enough $v_2$, to demonstrate that the measured data can be easily accommodated by models without invoking the CME.
\end{abstract}

\maketitle

%\begin{keyword}
%%% keywords here, in the form: keyword \sep keyword
%chiral magenetic effect\sep heavy ion collisions \sep quark gluon plasma \sep Hijing \sep AMPT 
%
%%25.75.-q 	Relativistic heavy-ion collisions (collisions induced by light ions studied to calibrate relativistic heavy-ion collisions should be classified under both 25.75.-q and sections 13 or 25 appropriate to the light ions)
%%25.75.Ag 	Global features in relativistic heavy ion collisions
%%25.75.Bh 	Hard scattering in relativistic heavy ion collisions
%%25.75.Cj 	Photon, lepton, and heavy quark production in relativistic heavy ion collisions
%%25.75.Dw 	Particle and resonance production
%%25.75.Gz 	Particle correlations and fluctuations
%%25.75.Ld 	Collective flow
%%25.75.Nq 	Quark deconfinement, quark-gluon plasma production, and phase transitions (see also 12.38.Mh Quark-gluon plasma in quantum chromodynamics; 21.65.Qr Quark matter in nuclear matter)
%\PACS 25.75.-q \sep 25.75.Gz \sep 25.75.Ld
%
%%% MSC codes here, in the form: \MSC code \sep code
%%% or \MSC[2008] code \sep code (2000 is the default)
%\end{keyword}

%\maketitle
%\linespread{1.6}

%%%%%%%%%%%%%%%%%%%%%%%%%%%%%%%%%%%%%%%%%%%%%%%%%%%%%%%%%%%%%%%
\section{Introduction}
In recent years the chiral magnetic effect (CME) in relativistic heavy ion collisions has attracted intense interests~\cite{Kharzeev:2015znc,Zhao:2018ixy,zhao:225,Zhao:2019hta}. The CME refers to an electric current along a strong magnetic field, produced in the early times of relativistic heavy ion collisions, perpendicular on average to the reaction plane (RP) of those collisions -- the plane spanned by the impact parameter direction and the beam~\cite{Kharzeev:2004ey,Kharzeev:2007jp,Fukushima:2008xe,Muller:2010jd,Liu:2011ys,Kharzeev:2013ffa}.
The electric current is a result of the motion of quarks in a metastable domain of imbalanced chirality, which can form from vacuum fluctuations in quantum chromodynamics (QCD)~\cite{Lee:1974ma,Kharzeev:1998kz,Kharzeev:1999cz}. Such an electric current of quarks results in a charge separation in the final state across the reaction plane.

Reaction-plane and charge-dependent correlations have been observed in relativistic heavy ion collisions, first by the STAR experiment at BNL's Relativistic Heavy Ion Collider (RHIC)~\cite{Abelev:2009ac,Abelev:2009ad,Adamczyk:2013hsi,Adamczyk:2014mzf} and later by experiments at the Large Hadron Collider (LHC)~\cite{Abelev:2012pa,Khachatryan:2016got,Sirunyan:2017quh,Acharya:2017fau}. 
Some, but not all, of the observed features are consistent with charge separation from the CME. 
Background correlations due to mundane physics were studied and it was initially concluded that no model studied could reproduce the measurements~\cite{Abelev:2009ac,Abelev:2009ad}. 
For example, the Hijing (Heavy ion jet interaction generator~\cite{Wang:1991hta,Wang:1996yf}) model that was studied (where the real RP was used) yielded far smaller magnitude for the opposite-sign charge correlation than observed in data~\cite{Abelev:2009ac,Abelev:2009ad}. 
This led to the conclusion that the CME had to be invoked to explain the observed correlations in heavy ion collisions~\cite{Abelev:2009ac,Abelev:2009ad}.

In this paper we show that this conclusion of the CME interpretation was premature. We show that the reason that Hijing predicts
a far smaller correlation than data is because the elliptic anisotrpy ($v_2$)~\cite{Reisdorf:1997fx} parameter in Hijing is too small. 
When properly scaled, the Hijing model can reproduce in entirety the measured correlations. 
We will also employ the AMPT (A multi-phase transport~\cite{Zhang:1999bd,Lin:2004en}) model, which has a large enough $v_2$, to demonstrate that the measured data can be easily accommodated by models without invoking the CME. 
Our studies reinforce the conclusion from other previous studies, contrary to that claimed in Refs~\cite{Abelev:2009ac,Abelev:2009ad,Kharzeev:2015znc}, 
that background correlations may account for all of the observed correlations~\cite{Wang:2009kd,Bzdak:2009fc,Schlichting:2010qia,Adamczyk:2013hsi,Wang:2016iov}.

\section{The Hijing and AMPT models}
In this study, we use two typical, commonly used models, namely the Hijing (v1.411) and AMPT (v2.26t5d6) to calculate charge correlations. 
Hijing is a QCD inspired model simulating heavy ion collisions by binary nucleon-nucleon (NN) collisions using the Glauber geometry, incorporating nuclear shadowing effects and energy loss of partons traversing the medium created in those collisions (jet quenching). 
It uses PYTHIA~\cite{Sjostrand:1985ys, Sjostrand:2006za} for generating kinematic variables of scattered partons for each hard or semihard interaction and Lund string fragmentation (JETSET) ~\cite{Andersson:1983ia} for hadronization. Jet quenching is included in our Hijing simulation.

We employ the string melting version of AMPT~\cite{Zhang:1999bd,Lin:2004en} in our study. 
The model consists of four components: the initial condition of collisions, partonic elastic scatterings, hadronization and hadronic scatterings. The initial condition in AMPT is provided by the Hijing model. 
The hadrons generated by Hijing are converted into valence quarks and antiquarks.
The subsequent parton-parton elastic scatterings are described by ZPC~\cite{Zhang:1997ej}. 
The Debye-screened differential cross-section $d\sigma/dt\propto\alpha_s^2/(t-\mu_D^2)^2$~\cite{Lin:2004en} is used for parton scattering. 
The strong coupling constant $\alpha_s=0.33$ and Debye screening mass $\mu_D=2.265$/fm are employed, so that the total parton-parton scattering cross section is $\sigma=3$~mb. 
After partons stop interacting, a simple quark coalescence model is applied to convert partons into hadrons~\cite{Lin:2004en,Lin:2014tya}. Subsequent interactions of those formed hadrons are modeled by ART ~\cite{Li:1995pra}. Hadronic interactions include meson-meson, meson-baryon, and baryon-baryon elastic and inelastic scatterings. More details can be found in Ref.~\cite{Lin:2004en}.

\section{The $\Delta\gamma$ correlator}
The common observable to study the charge-dependent and reaction-plane-dependent azimuthal correlations is the $\Delta\gamma$ variable~\cite{Voloshin:2004vk}. It is the difference of the opposite-sign (OS) and same-sign (SS) correlators,
\begin{eqnarray}
 \Delta\gamma = \gamma_{OS} - \gamma_{SS} \,,
\label{eq:0}
 \end{eqnarray}
such that one of the main physics backgrounds, the momentum conservation, is cancelled~\cite{Abelev:2009ac,Abelev:2009ad}.
The correlators are defined by
\begin{eqnarray}
 \gamma_{\alpha\beta} = \langle\cos(\phi_\alpha+\phi_\beta-2\psi)\rangle \,,
\label{eq:1}
\end{eqnarray}
where $\phi_\alpha$ and $\phi_\beta$ are the azimuthal angles of two particles, either OS or SS, and $\psi$ is the azimuthal angle of the reaction plane. The reaction plane is not measured, and is approximated by the event plane reconstructed from particle momenta in the final state. The inaccuracy is corrected by the event plane resolution. The event plane can also be taken as the direction of a single particle, called particle $c$. The resolution is simply given by the elliptic flow parameter of particle $c$, $v_{2,c}$. This is called the three-particle method~\cite{Voloshin:2004vk,Abelev:2009ac,Abelev:2009ad}:
\begin{eqnarray}
 \gamma_{\alpha\beta} = \langle\cos(\phi_\alpha+\phi_\beta-2\phi_c)\rangle/v_{2,c} \,,
\end{eqnarray}
Physics backgrounds arise when particles $\alpha$ and $\beta$ are intrinsically correlated, not due to a global flow correlation to a common plane~\cite{Voloshin:2004vk,Wang:2009kd,Wang:2016iov}. 
The intrinsic correlation is sometimes dubbed as nonflow correlation. One example is nonflow correlations due to resonance decays, primarily affecting OS correlations. In such a case, the background can be expressed as
\begin{eqnarray}
\Delta\gamma_{\rm reso.} 
&=& \langle\cos(\phi_\alpha+\phi_\beta-2\phi_{\rm reso.})\rangle \cdot v_{2,{\rm reso.}} \nonumber\\
&=& \frac{N_{\rm reso.}}{N_\alpha N_\beta} \langle\cos(\phi_\alpha+\phi_\beta-2\phi_{\rm reso.})\rangle \cdot v_{2,{\rm reso.}} \,,
\label{eq:bkgd}
\end{eqnarray}
where $v_{2,{\rm reso.}}= \langle\cos2(\phi_{\rm reso.}-\psi)\rangle$ is the resonance elliptic flow parameter. In the above equation, $\phi_\alpha$ and $\phi_\beta$ are the azimuthal angles of the two decay daughters, so the quantity $\langle\cos(\phi_\alpha+\phi_\beta-2\phi_{\rm reso.})\rangle$ is determined by decay kinematics, insensitive to collision centralities or types.
Other nonflow background correlations include (mini)jets~\cite{Petersen:2010di}, or more generally, cluster correlations~\cite{Wang:2009kd}.

\section{Results and Discussions}
\begin{figure}[phbt] %[here,bottom,top]
  \begin{center}
    \includegraphics[width=0.7\textwidth]{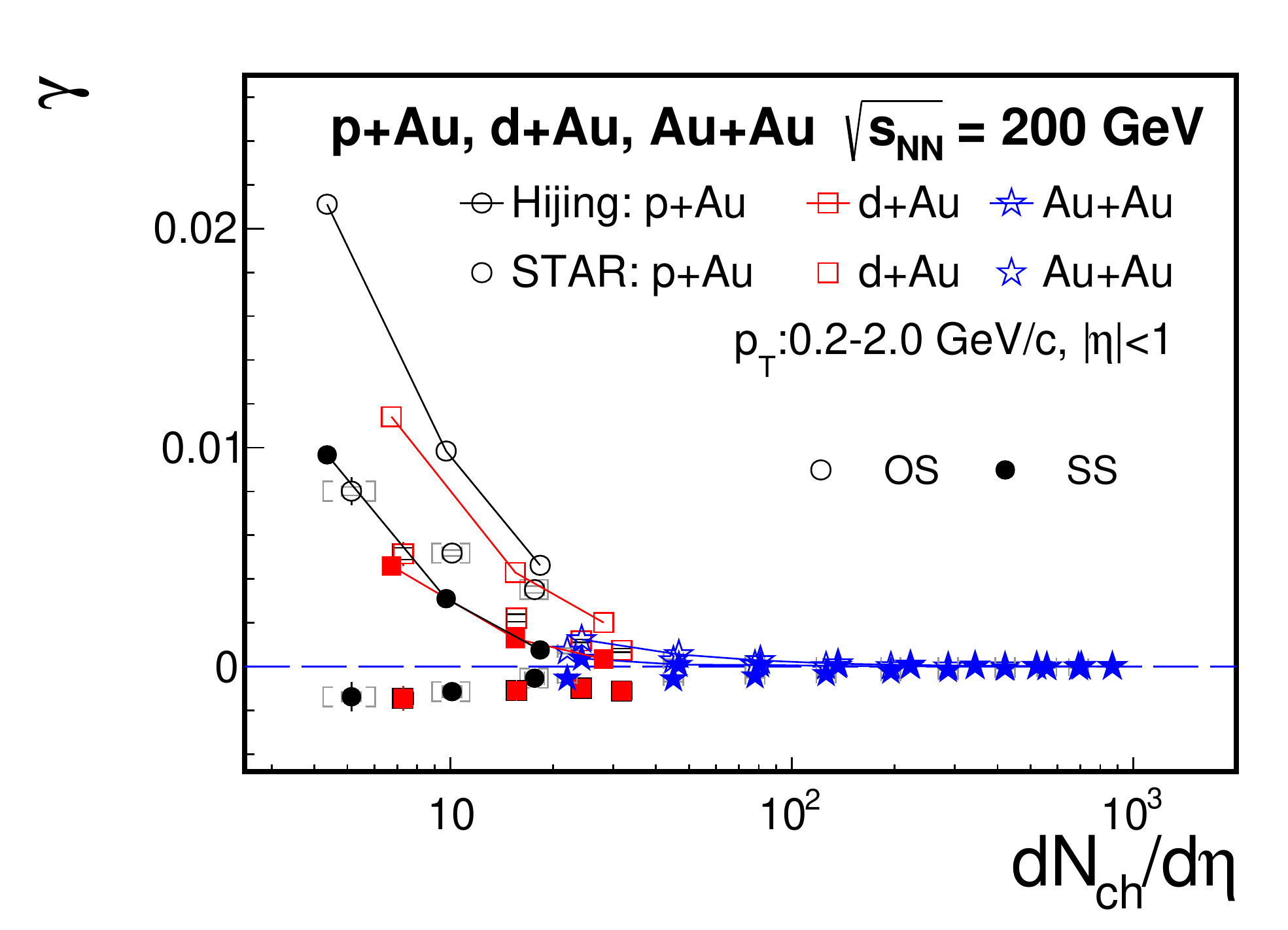}
    \caption{Hijing predictions of the opposite-sign (OS, open symbols) and same-sign (SS, filled symbols) $\gamma$ correlators, with comparisons to data~\cite{Abelev:2009ac,Abelev:2009ad,STAR:2019xzd}. The model predictions are connected by lines. The data symbols are same as the corresponding model symbols, but not connected by lines. The \pAu\ results are shown in circles, \dAu\ in squares, and Au+Au in stars. The results are plotted as functions of the mid-rapidity charged hadron multiplicity density, $\dndeta$.}
    \label{fig:g}
  \end{center}
\end{figure}

Figure~\ref{fig:g} shows the $\gOS$ and $\gSS$ correlators in Hijing compared to experimental data~\cite{Abelev:2009ac,Abelev:2009ad,STAR:2019xzd}. The model data are binned in multiplicity similarly to experimental data to correspond to the cross-section fractions. The charged hadron multiplicity within pseudorapidity $-0.5<\eta<0.5$ is used. 
The multiplicity cut values are not the same between the models and the data, because the models do not exactly reproduce the data multiplicity and because of the detection inefficiencies in data that are not included in the models. 
The model and data results are plotted against the midrapidity charged hadron pseudorapidity density, $\dndeta$. The decreasing $\gamma$ amplitudes with increasing $\dndeta$ is mainly due to the trivial multiplicity dilution effect. 
The $\gOS$ values from Hijing have the same sign as the experimental data. The $\gSS$ values are more different from the experimental data; $\gSS$ from Hijing are mostly positive, while the experimental data are mostly negative.
The discrepancy between Hijing results and experimental data are mostly from the charge independent background, such as the momentum conservation effect~\cite{Schlichting:2010qia,Bzdak:2010fd,Pratt:2010zn}.

\begin{figure}[phbt] %[here,bottom,top]
  \begin{center}
    \includegraphics[width=0.7\textwidth]{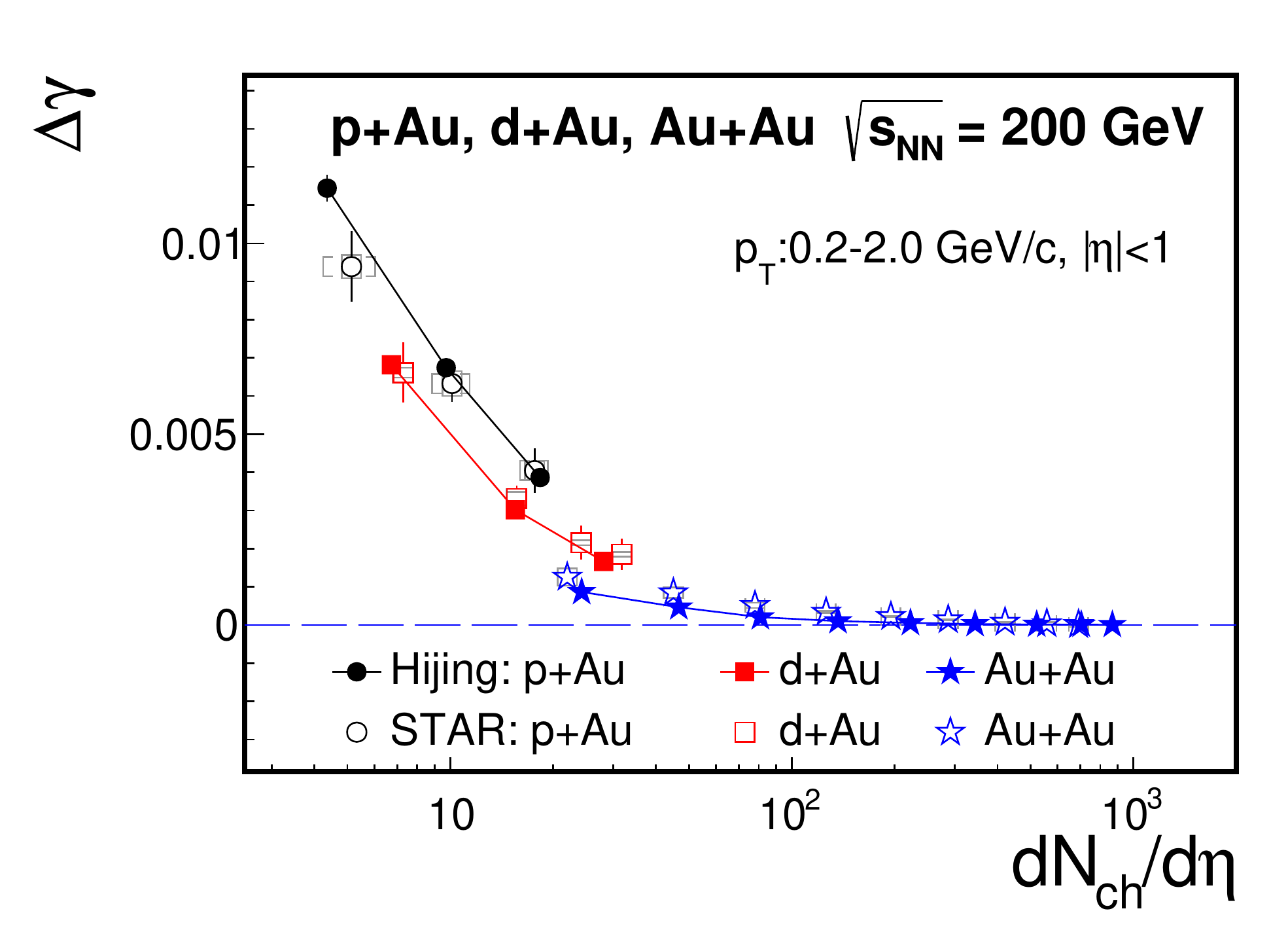}
	  \caption{Hijing predictions of the $\dg$ correlator (filled symbols), with comparisons to data (open symbols)~\cite{Abelev:2009ac,Abelev:2009ad,STAR:2019xzd}. The model predictions are connected by lines. The \pAu\ results are shown in circles, \dAu\ in squares, and Au+Au in stars. The results are plotted as functions of the mid-rapidity charged hadron multiplicity density, $\dndeta$.} 
    \label{fig:dg}
  \end{center}
\end{figure}

\begin{figure}[phbt] %[here,bottom,top]
  \begin{center}
    \includegraphics[width=0.7\textwidth]{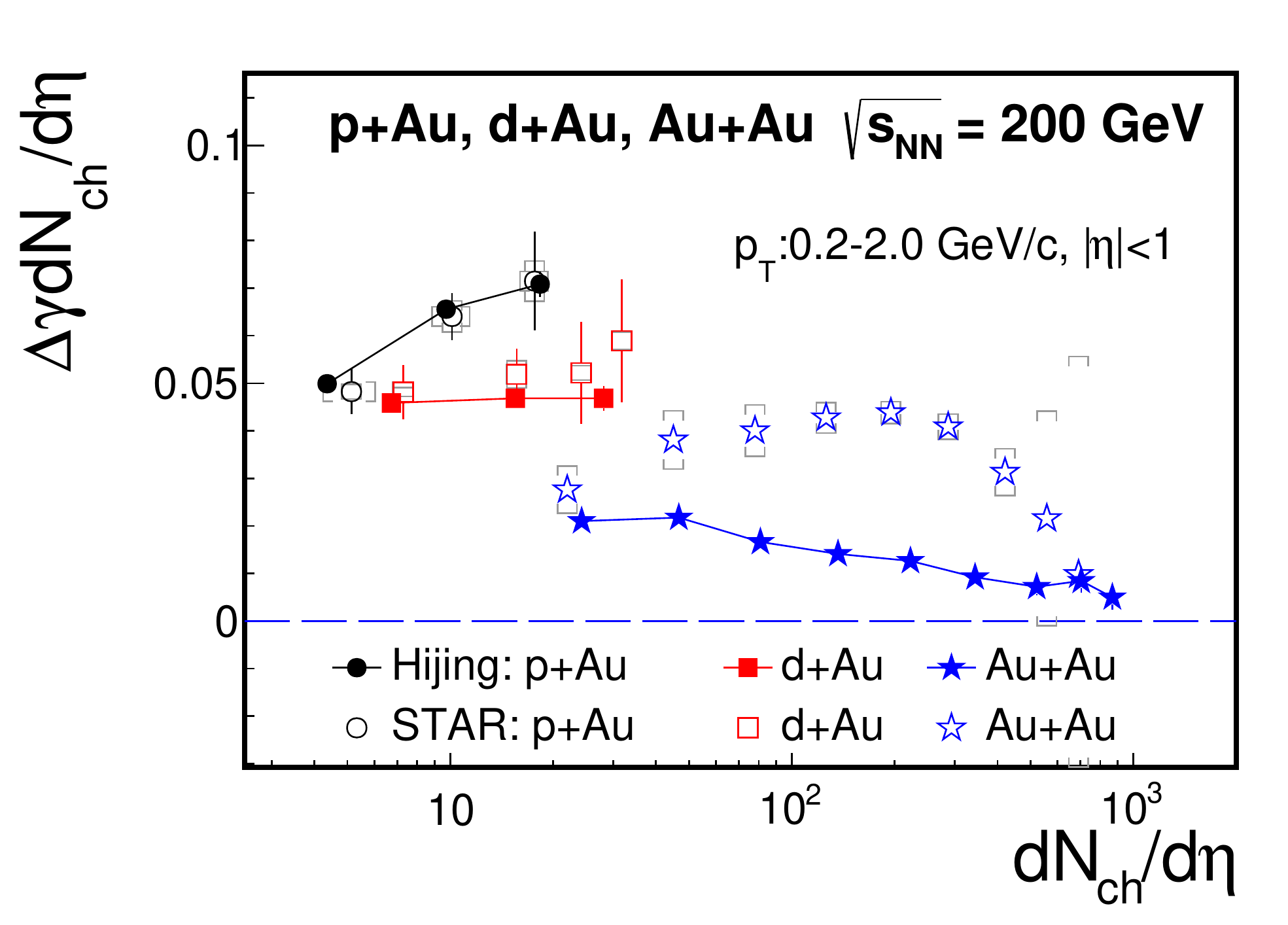}
	  \caption{
		  Hijing predictions of the multiplicity scaled correlator (filled symbols), with comparisons to data (open symbols)~\cite{Abelev:2009ac,Abelev:2009ad,STAR:2019xzd}. The model predictions are connected by lines. The \pAu\ results are shown in circles, \dAu\ in squares, and Au+Au in stars. The results are plotted as functions of the mid-rapidity charged hadron multiplicity density, $\dndeta$.
	  }
    \label{fig:dgN}
  \end{center}
\end{figure}

To eliminate the charge-independent background sources, Fig.~\ref{fig:dg} shows the $\dg$ correlator in Hijing. 
Further to remove the trivial multiplicity dilution effect and to better show the heavy ion data, 
the $\dg$ is multiplied by $\dndeta$ in Fig.~\ref{fig:dgN}. 
The Hijing results are compared to experimental data~\cite{Abelev:2009ac,Abelev:2009ad,STAR:2019xzd}. 
The Hijing results agree well with the data in small system \pAu\ and \dAu\ collisions. 
On the other hand, the Hijing results in Au+Au collisions are much smaller than the data. This has been interpreted as a supporting evidence for the possible CME in experimental data~\cite{Abelev:2009ac,Abelev:2009ad}. 

\begin{figure}[phbt] %[here,bottom,top]
  \begin{center}
    \includegraphics[width=0.7\textwidth]{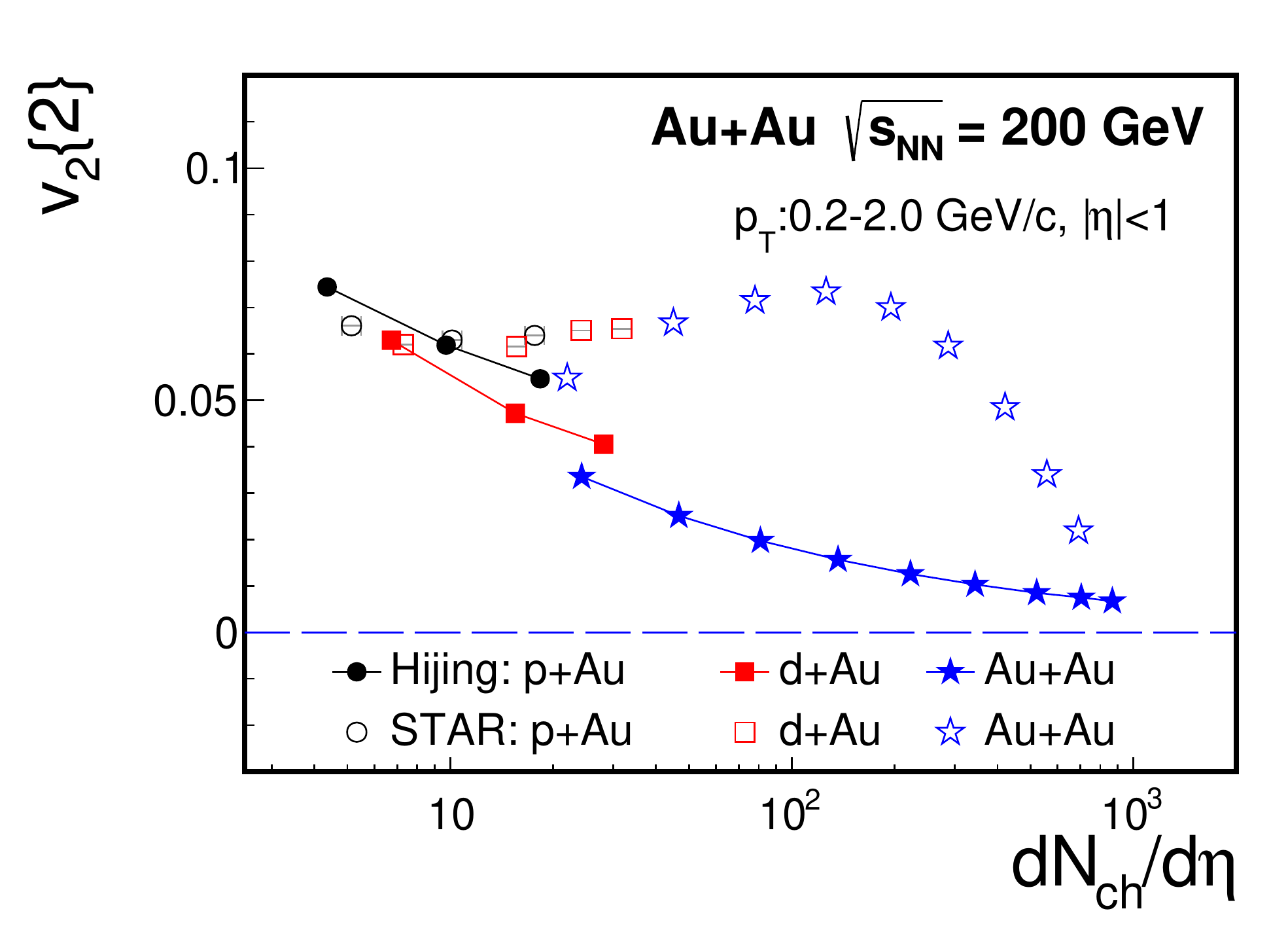}
    \caption{
		Hijing predictions of the $v_{2}$ parameter (filled symbols), with comparisons to data (open symbols)~\cite{STAR:2019xzd}. The model predictions are connected by lines. The \pAu\ results are shown in circles, \dAu\ in squares, and Au+Au in stars. The results are plotted as functions of the mid-rapidity charged hadron multiplicity density, $\dndeta$.
	  }
    \label{fig:v2}
  \end{center}
\end{figure}

The $\dg$ signal in Hijing is due to background correlations. According to Eq.~(\ref{eq:bkgd}), the background correlations are proportional to $v_2$. In Fig.~\ref{fig:v2} we show the $v_2$ parameters from Hijing and compare them to those from experimental data~\cite{Abelev:2009ac,Abelev:2009ad,STAR:2019xzd}. 
Indeed, the $v_2$ values are similar between Hijing and data in small-system collisions, so the $\dg$ values are also similar. In heavy ion collisions, however, the $v_2$ values in Hijing are much smaller than data.
Similarly, as shown in Fig.~\ref{fig:dgN}, the $\dg$ values in Hijing are much smaller than data too.

\begin{figure}[phbt] %[here,bottom,top]
  \begin{center}
    \includegraphics[width=0.6\textwidth]{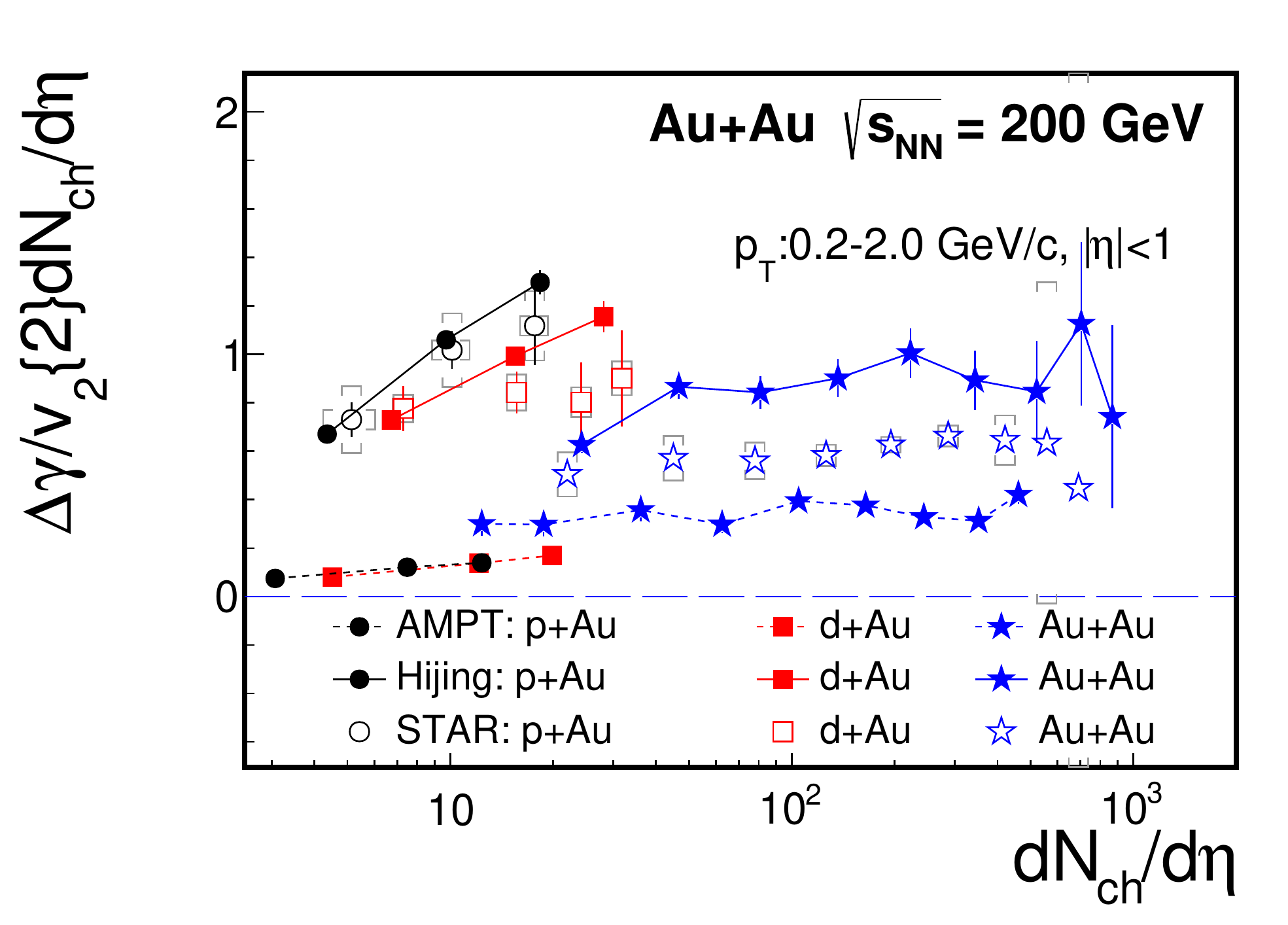}
    \caption{Hijing and AMPT predictions of the scaled correlator, $\dndeta\cdot\dg/v_2$ (filled symbols), with comparisons to data (open symbols)~\cite{STAR:2019xzd}. 
	The Hijing predictions are connected by solid lines and the AMPT results are connected by dashed lines. The \pAu\ results are shown in circles, \dAu\ in squares, and Au+Au in stars. The results are plotted as functions of the mid-rapidity charged hadron multiplicity density, $\dndeta$
	  }
    \label{fig:dgscale}
  \end{center}
\end{figure}

According to Eq.~(\ref{eq:bkgd}), the source of the background is the correlation term, $\langle\cos(\phi_\alpha+\phi_\beta-2\phi_{\rm reso.})\rangle$. This term reflects the intrinsic correlation between particles, such as the daughter particles from a resonance decay. In the background picture, therefore, the more direct quantity is the scaled correlator
\begin{equation}
	\dg_{\rm scaled}=\dndeta\cdot\dg/v_2\propto \frac{dN_{\rm reso.}/d\eta}{\dndeta}  \langle\cos(\phi_\alpha+\phi_\beta-2\phi_{\rm reso.})\rangle.
\label{eq:dgscale}
\end{equation}
Figure~\ref{fig:dgscale} shows the scaled $\dgscale$ correlator from Hijing, compared to data. Now there is not much difference between Hijing and data, unlike those shown in Fig.~\ref{fig:dg} and Fig.~\ref{fig:dgscale}. Furthermore, there is not much difference overall in this quantity between small systems and big systems. 
This makes sense because the intrinsic particle correlations reflect the underlying physics mechanisms
for the correlations, such as the decay kinematics, and should not be very different between different systems.

Also shown in Fig.~\ref{fig:dgscale} are the corresponding results from AMPT.
In heavy-ion collisions the scaled $\dgscale$ correlator in AMPT is also similar to data. 
The $v_2$ in AMPT, in contrast to Hijing, is known to reproduce data well~\cite{Lin:2014tya}. 

Quantitatively, however, the models do not reproduce the data. The Hijing $\dg_{\rm scaled}$ overpredicts Au+Au data,
whereas the AMPT underpredicts the Au+Au data by a similar amount.
The Hijing seems to well reproduce the small system data, but AMPT predicts a significantly weaker magnitude.
These discrepancies could arise from a number of reasons.
(1) Hadronic rescatterings can destroy resonances, and this could be a reason why the Au+Au data are lower than Hijing which does not include hadronic rescatterings. AMPT could have too many rescatterings resulting in weaker correlations. It is also possible that 
the reason is due to the lack of minijet correlations or that too few resonances are included in AMPT.
On the other hand, hadronic rescattering would yield a decreasing correlation with increasing centrality, which is at odd with the Au+Au results in Fig.~\ref{fig:dgscale}, but there could be other effects compensating a decreasing trend.
(2)The fact that Hijing reproduces the small system data well may indicate that the minijet correlations
are modeled well by Hijing. The Hijing results keep increasing with $dN_{ch}/d\eta$ in small systems, and this could be due to increasing jet correlations biased by the requirement of the high multiplicities~\cite{Adamczyk:2014fcx}. The increase in the data is not as significant, perhaps due to the rescattering effect aforementioned.
(3) The AMPT results in small systems are a factor of several lower than the data. This is likely due to the fact that minijet correlations are destroyed in the AMPT initialization using Hijing output.
It is unclear why the overall correlation strengths differ by a factor of 2 or so between small systems and heavy ion collisions in AMPT, unlike Hijing. Further investigation is needed.

Note that the backgrounds arise from correlations of the background sources 
with the reconstructed event plane or the third particle $c$, and thus are propagated into the three-particle correlator. 
The physics nature of the correlations with the event plane or the particle $c$ is unimportant for the background explanation of the $\dg$ correlator. 
For example, the correlation to the event plane or $c$ in Hijing is likely due to jets (e.g.~a resonance and the particle $c$ are parts of a dijet) or multiparticle clusters (e.g.~from string decays); the correlation to event plane or $c$ in AMPT is likely due to collective elliptic flow,
at least for heavy ion collisions,
such that almost all particles of the event are correlated.

\section{Summary}
The background contamination in the CME-sensitive $\dg$ observable arises from intrinsic particle correlations (nonflow). Those nonflow correlations include resonance decays, clusters of multiparticle correlations, and (mini)jets. 
We employed the Hijing and AMPT models to study the effect of those backgrounds. 
Hijing seems to contain similar strength of those backgrounds as in data.
Because of the weaker correlation to event plane or $c$, the final $\dg$ observable in Hijing is much smaller than the heavy ion data. If the collective flow was present in Hijing, then the data would be well reproduced as indicated by the comparisons of the scaled $\dgscale$ correlator. 
AMPT, on the other hand, does not seem to contain enough correlations as in data as indicated by the small system results. 
This could be due to the fact that minijets are not included in AMPT, not all high mass resonances are included, and/or resonance decay daughters rescatter and lose their correlations from decay. As a result, although AMPT has enough $v_2$, the $\dg$ values in AMPT are underpredicted. 

The models do not necessarily reproduce data exactly. 
However, one cannot conclude that there must be CME in the heavy ion data just because the data $\dg$ is larger than that in models. 
In the case of AMPT, this would not explain the small system results where any CME would be small, yet AMPT is off from data by a large amount. 
The reason Hijing does not reproduce data in terms of $\dg$ is because Hijing does not have enough $v_{2}$.
There have been claims that the CME had to be invoked because no model studied, including Hijing, could reproduce data. This conclusion was premature as we have demonstrated in this work. 

The physics backgrounds are dominant in the CME-sensitive $\Delta\gamma$ observable. When backgrounds dominate, one should be careful not to overly rely on models. 
Models in this case are useful only to guide one's thinking, but cannot be used for quantitative predictions of the backgrounds. 
This is because a small deviation of the model from reality could give a 
large error on the extracted signal from data treating the model as background, potentially
leading to a wrong conclusion. 
Backgrounds have to be rigorously subtracted by data-driven methods before any conclusion about the CME can be made~\cite{Sirunyan:2017quh,Acharya:2017fau,Xu:2017qfs,Zhao:2017nfq}.

\section*{Acknowledgments} %no section number by adding '*'
We thank Dr.~Wei Li for fruitful discussions. This work is supported in part by US~Department of Energy Grant No.~DE-SC0012910, the National Natural Science Foundation of China under Grant No.~11847315, 11947410 and the Natural Science Foundation of Hubei Province under Grant No.~2019CFB563. 

\bibliographystyle{unsrt} %include paper titles
\bibliography{ref} %include your ref's in ref.bib
\end{document}